\documentclass[cameraready]{Interspeech}

\title{Speaker Disentanglement of Speech Pre-trained Models via Interpretability}

\author[affiliation={2}]{Junhua}{Li}
\author[affiliation={1}]{Xiaoxu}{Zhu}
\author[affiliation={3}]{Aaron J.}{Li}

\address{
    $^1$ XPENG Motors, Beijing, China \\
    $^2$ China National Aviation Fuel Group Limited, Beijing, China \\
    $^3$ University of California, Berkeley, CA, USA
}

\email{}


\usepackage{subcaption}
\usepackage{multirow}
\usepackage{booktabs}

\begin{document}

\maketitle

\begin{abstract}
Self-supervised speech models entangle content and speaker information, causing speaker
bias in downstream tasks and enabling identity leakage from anonymized representations.
We make two contributions. InterpTRQE-SptME is a benchmark that directly quantifies
residual speaker information in content embeddings using SHAP-based interpretability,
providing a direct measure of speaker entanglement---unlike existing metrics that
rely solely on downstream task performance. InterpTF-SptME exploits these SHAP insights
to suppress dimensions that encode speaker identity without model retraining. Across
LibriTTS and VCTK with 40 English speakers and seven models, including HuBERT, WavLM,
and ContentVec, SHAP Noise filtering reduces speaker residuals from 18.82\% to near zero
while keeping the word error rate at or below 2.3\%. The method is
model-agnostic and label-free.
\end{abstract}

\section{Introduction}

Recent self-supervised speech models—wav2vec~2.0~\cite{baevski2020wav2vec2},
HuBERT~\cite{hsu2021hubert}, and WavLM~\cite{chen2022wavlm}—learn rich hierarchical
representations that serve a wide range of downstream tasks. Yet the same representations
that make these models versatile also entangle two fundamentally distinct attributes:
phonetic content and speaker identity. This entanglement creates two interrelated
problems. For content-centric tasks such as automatic speech recognition (ASR) and voice
conversion, residual speaker information introduces systematic bias that degrades
performance. For privacy-sensitive deployments, speaker identity leaks through
representations that are nominally anonymized, undermining efforts like the VoicePrivacy
initiative~\cite{tomashenko2020introducing}.

Existing solutions, most notably ContentVec~\cite{qian2022contentvec}, tackle
disentanglement through dedicated training pipelines that combine voice conversion,
contrastive loss, and speaker conditioning. While effective, such approaches require
substantial compute and architectural commitment, and they still provide no direct
measurement of how much speaker information remains after disentanglement. Indirect
proxies—speaker identification accuracy, phone purity, or ASR word error rate—conflate
multiple factors and cannot isolate the residual speaker contribution in a
representation.

We address both gaps through interpretability. SHAP (SHapley Additive
exPlanations)~\cite{lundberg2017unified} computes each feature dimension's marginal
contribution to a model's output. By applying gradient SHAP to a speaker classifier
whose input concatenates content embeddings with a reference speaker embedding, we obtain
a principled, dimension-wise attribution of speaker information within content
representations. This attribution simultaneously gives a direct measurement (our
InterpTRQE-SptME benchmark) and a filtering recipe (our InterpTF-SptME method).

The main contributions of this work are: (1)~InterpTRQE-SptME, the first benchmark for
directly quantifying speaker residuals in content embeddings, applicable to any
pre-trained model without retraining; (2)~InterpTF-SptME, a post-hoc filtering method
with two strategies—SHAP Noise and SHAP Cropping—that reduce speaker residuals from
18.82\% to near zero with minimal impact on ASR quality; and (3)~systematic evaluation
across seven models and 40 English speakers drawn from LibriTTS~\cite{zen2019libritts}
and VCTK~\cite{veaux2017cstr}, revealing layer-wise speaker accumulation patterns and
outperforming ContentVec in disentanglement without any retraining.

\section{Related Work}

\subsection{Speaker disentanglement in speech representations}

Separating speaker identity from linguistic content has been approached from multiple
directions. GAN-based methods such as StarGAN-VC~\cite{kameoka2018stargan} and
CycleGAN-VC~\cite{kaneko2018cyclegan} train discriminators to push generators toward
speaker-invariant codes, while VAE-based systems including AutoVC~\cite{qian2019autovc}
and VoiceMixer~\cite{lee2021voicemixer} exploit information bottlenecks to limit what
the content encoder can retain. These models demonstrate that controlled disentanglement
is feasible, yet they are purpose-built architectures trained end-to-end for voice
conversion and cannot be applied to an existing pre-trained model without full
retraining.

Self-supervised models offer stronger content representations but inherit the
entanglement problem. Discrete speech units obtained by clustering HuBERT features
effectively remove speaker cues but sacrifice linguistic detail, increasing pronunciation
errors~\cite{niekerk2022comparison}. Soft unit representations alleviate the precision
loss~\cite{champion2022disentangled} but provide no direct measure of remaining speaker
information. ContentVec~\cite{qian2022contentvec} is a direct SSL-based
disentanglement effort: it uses voice conversion to remove speaker information from
teacher labels, adds contrastive loss over speaker-augmented views, and conditions
separately on speaker embeddings. The resulting model reduces speaker identification
accuracy from 73.7\% to 37.7\% on VCTK---yet this indirect measure does not quantify
how much speaker information actually remains in the embedding space. Our benchmark fills
this gap, and our filtering method shows that post-hoc SHAP-guided perturbation of
HuBERT LARGE embeddings surpasses ContentVec's residual level without any retraining.

Speaker anonymization presents a related privacy challenge addressed by the VoicePrivacy
initiative~\cite{tomashenko2020introducing}. Anonymization systems typically modify
speaker embeddings explicitly; our work instead targets the content embedding pathway,
which is generally expected to be speaker-free but demonstrably is not.

\subsection{Interpretability for speech models}

Interpretability methods have advanced rapidly in NLP and computer vision but remain
underexplored in speech. SHAP~\cite{lundberg2017unified} and LIME~\cite{ribeiro2016lime}
provide post-hoc, model-agnostic feature attributions. In the speech domain, LIME
variants have been applied to phoneme recognition in Kaldi/TIMIT
systems~\cite{wu2023can,mazumder2023harnessing}. Layer-wise relevance propagation (LRP)
has been integrated into GRU-based ASR models to identify sample-level relevance scores
while preserving accuracy~\cite{bharadhwaj2018layer}. Gradient-based saliency methods
have been used to probe which acoustic features drive ASR errors. Despite this progress,
no prior work has applied interpretability to directly measure or mitigate speaker
entanglement in pre-trained speech representations.

Several studies in NLP demonstrate that explanations can go beyond passive diagnosis to
actively improve model behavior. SHAP values used as loss weights improve radar-based
people counting by 4\%~\cite{sharma2022utilizing}, and the AMPLIFY
framework~\cite{krishna2023post} applies post-hoc corrections to LLM predictions.
Translating this principle to speech representations motivates our filtering design:
SHAP attributions identify which dimensions encode speaker identity, and we use that
knowledge to perturb those dimensions selectively.

\section{Methodology}

\subsection{InterpTRQE-SptME benchmark}
\label{ssec:benchmark}

The benchmark provides a direct, model-agnostic quantification of speaker
information residuals in content embeddings. Given a speech sample $x$, we extract two
representations: a content embedding from model layer $l$,
\begin{equation}
  E_c = M^{(l)}(x) \in \mathbb{R}^{T \times d_c},
\end{equation}
time-averaged to $\bar{E}_c \in \mathbb{R}^{d_c}$ ($T$ frames, $d_c \in \{768, 1024\}$
depending on model), and a speaker embedding from a pre-trained
ECAPA-TDNN~\cite{desplanques2020ecapa} via SpeechBrain~\cite{ravanelli2021speech},
\begin{equation}
  E_s = S(x) \in \mathbb{R}^{d_s}, \quad d_s = 192.
\end{equation}

The two embeddings are concatenated and passed to a 4-layer MLP speaker classifier:
\begin{equation}
  E_{\mathrm{cat}} = [\bar{E}_c;\, E_s] \in \mathbb{R}^{d_c + d_s}, \qquad
  \hat{y} = f(E_{\mathrm{cat}}).
\end{equation}
Because $E_{\mathrm{cat}}$ explicitly contains $E_s$, the classifier can be trained to
100\% training accuracy. Gradient SHAP~\cite{lundberg2017unified} then attributes each
input dimension's
contribution $\phi_i$ to the speaker classification decision. Since both content and
speaker segments are included in the input, the ratio
\begin{equation}
  R_M = \frac{\sum_{i=1}^{d_c} |\phi_i|}{\sum_{i=1}^{d_c} |\phi_i| +
        \sum_{j=1}^{d_s} |\phi_{d_c+j}|}
  \label{eq:residual}
\end{equation}
measures what fraction of the classifier's evidence comes from content dimensions.
$R_M = 0$ would indicate perfect disentanglement (all speaker information is carried by
$E_s$, none by $E_c$), while high $R_M$ signals substantial speaker leakage into the
content embedding. The speaker embedding $E_s$ serves as a fixed, cross-model reference
baseline, enabling fair comparison across architectures with different hidden dimensions.

\subsection{InterpTF-SptME: interpretability-guided filtering}
\label{ssec:filter}

Given SHAP attribution values $\phi_c \in \mathbb{R}^{d_c}$ for content dimensions, we
design two complementary filters that translate these attributions into targeted
perturbations of the content embedding $E_c \in \mathbb{R}^{T \times d_c}$.

\smallskip\noindent\textbf{SHAP Noise.} Attributions are first standardized to obtain a
zero-mean, unit-variance profile over embedding dimensions:
\begin{equation}
  \hat{\phi}_c = \frac{\phi_c - \mu_{\phi_c}}{\sigma_{\phi_c}}.
  \label{eq:noise_phi}
\end{equation}
Here $\hat{\phi}_c$ encodes the \emph{relative} speaker contribution of each dimension.
Dimensions with large positive $\hat{\phi}_c$ strongly support speaker classification;
those with large negative values suppress it; near-zero dimensions are content-dominant.
A structured noise vector is then constructed as
\begin{equation}
  n_{\mathrm{shap}} = \hat{\phi}_c \cdot \epsilon \cdot |\sigma| + \mu,
  \label{eq:noise_n}
\end{equation}
where $\epsilon \sim \mathcal{N}(0,I)$ introduces stochasticity, $|\sigma|$ controls
the perturbation magnitude (the negative convention reflects the suppressive intent: larger
$|\sigma|$ applies stronger noise to speaker-encoding dimensions), and $\mu = 0$
prevents systematic drift. The filtered embedding is
\begin{equation}
  E'_c = E_c + n_{\mathrm{shap}},
  \label{eq:noise_filter}
\end{equation}
where $n_{\mathrm{shap}}$ is broadcast across the $T$ frame dimension. Content
dimensions with near-zero $\hat{\phi}_c$ experience negligible perturbation, preserving
linguistic information.

\smallskip\noindent\textbf{SHAP Cropping.} An alternative strategy selectively suppresses the top-$r$\%
dimensions by absolute SHAP value via a soft binary mask:
\begin{equation}
  m_i = \begin{cases}
    0, & \phi_i \in \mathrm{top}\text{-}r\%  \\
    1, & \text{otherwise,}
  \end{cases}
  \label{eq:crop_mask}
\end{equation}
where $m_i = 0$ flags a dimension as a suppression candidate. The filtered embedding is
\begin{equation}
  E'_c = E_c \odot \bigl(m + \alpha\cdot(1-m)\bigr),
  \label{eq:crop_filter}
\end{equation}
with $\alpha \in (0,1)$ acting as a soft-cut coefficient: suppressed dimensions are
scaled to $\alpha$ of their original magnitude rather than zeroed out completely, which
preserves information pathways and avoids catastrophic content degradation.

\begin{figure*}[t]
  \centering
  \includegraphics[width=0.85\textwidth]{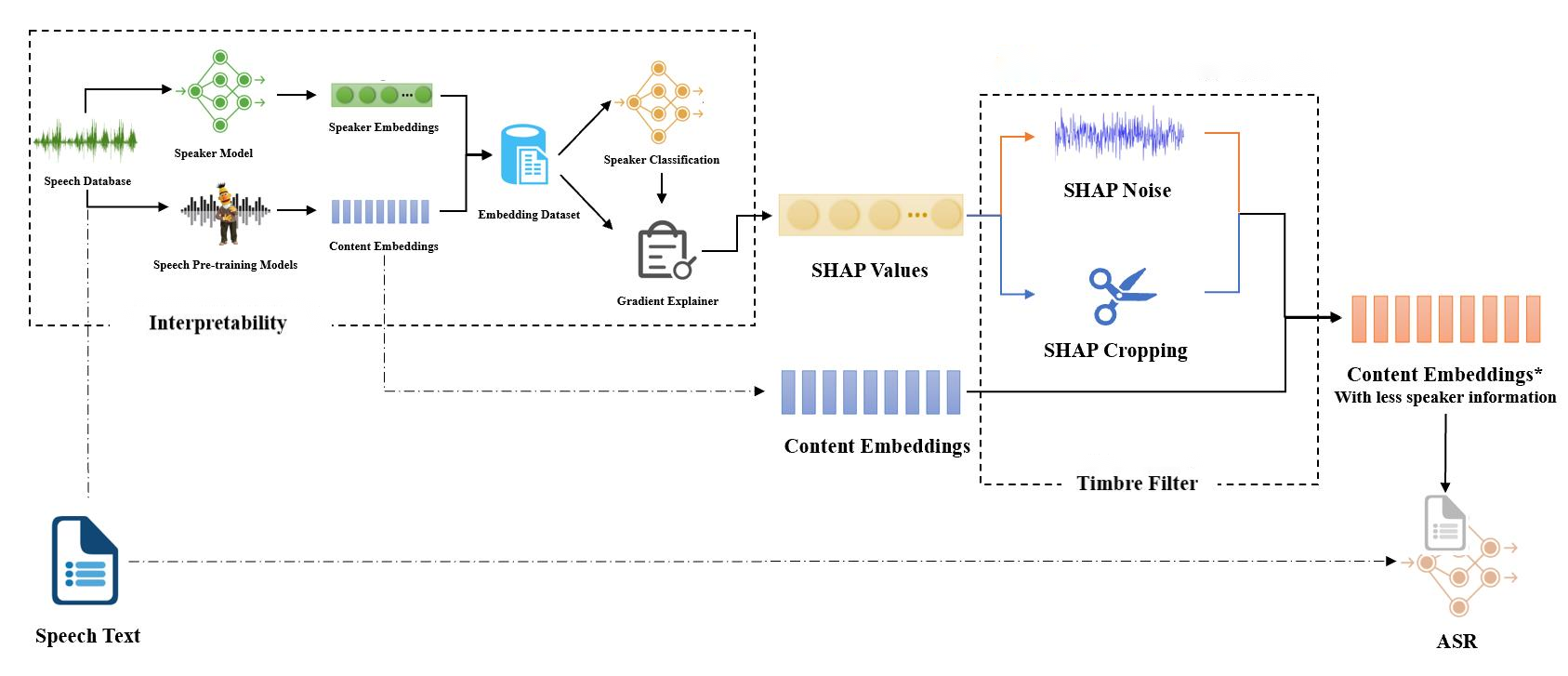}
  \caption{Overview of the proposed framework. Left: the InterpTRQE-SptME benchmark
    pipeline---content embeddings and reference speaker embeddings are concatenated,
    classified, and explained via gradient SHAP to obtain the timbre residual ratio
    $R_M$. Right: InterpTF-SptME filtering---SHAP attribution profiles guide SHAP Noise
    or SHAP Cropping transforms applied to content embeddings before downstream use.}
  \label{fig:framework}
\end{figure*}

\section{Experiments}

\subsection{Experimental setup}
\label{ssec:setup}

\smallskip\noindent\textbf{Data.} We evaluate on a combined corpus of 40 English speakers drawn from
LibriTTS~\cite{zen2019libritts} (train-clean-100 split, 20 speakers) and
VCTK~\cite{veaux2017cstr} (first 20 speakers, p225--p246 excluding p235 and p242). All audio is resampled to 16\,kHz, 16-bit
mono WAV and peak-normalized to $-3$\,dB. This combination provides diversity across
accents and recording conditions.

\smallskip\noindent\textbf{Models.} Seven speech pre-trained models are evaluated spanning a range of
architectures and training objectives: HuBERT BASE~\cite{hsu2021hubert} (layer~9),
HuBERT LARGE~\cite{hsu2021hubert} (layers~18 and 21), DPHuBERT~\cite{peng2023dp}
(layer~12), ContentVec~\cite{qian2022contentvec} (layer~12), WavLM
Base+~\cite{chen2022wavlm} (layer~12), Whisper-ppg (encoder output), and HuBERT-CH (a
Chinese fine-tuned variant, layer~9).

\smallskip\noindent\textbf{Benchmark implementation.} SHAP analysis uses Captum's GradientShap with 256
randomly selected baseline samples. The speaker classifier is a 4-layer MLP
(input$\,{\to}\,2048\,{\to}\,1256\,{\to}\,64\,{\to}\, N_{\text{spk}}$) trained with Adam
($\mathrm{lr}=10^{-4}$, batch 32) for 50 epochs to 100\% training accuracy. Local
smoothing ($\sigma=0.1$) is applied during SHAP computation to reduce single-sample
noise.

\smallskip\noindent\textbf{Content preservation metric.} Word Error Rate (WER) is computed using
seed-tts-eval~\cite{anastassiou2024seedtts} on outputs decoded from the
HuBERT-Large-Finetuned ASR model (facebook/hubert-large-ls960-ft, fine-tuned on
LibriSpeech 960h~\cite{panayotov2015librispeech}). Filtered embeddings are injected into
the ASR model via forward hooks at layer~21, replacing the pre-trained model's outputs
before decoding. CTC loss is additionally reported in Figure~\ref{fig:param_analysis} as
a granular local diagnostic.

\subsection{Benchmark results}
\label{ssec:benchmark_results}

Table~\ref{tab:benchmark} reports timbre residual across all seven models on the
combined 40-speaker corpus.

\begin{table}[th]
  \caption{Timbre residual $R_M$ (\%) across seven speech pre-trained models (40
    English speakers, LibriTTS+VCTK). Lower is better.}
  \label{tab:benchmark}
  \centering
  \begin{tabular}{lc}
    \toprule
    \textbf{Model} & \textbf{Residual (\%)} \\
    \midrule
    HuBERT BASE (L9)       & 14.23 \\
    HuBERT LARGE (L18)     & 11.02 \\
    HuBERT LARGE (L21)     & 18.82 \\
    DPHuBERT (L12)         & 7.98  \\
    ContentVec (L12)       & \textbf{5.47} \\
    WavLM Base+ (L12)      & 9.31  \\
    Whisper-ppg (Enc)      & 7.74  \\
    HuBERT-CH (L9)         & 14.15 \\
    \bottomrule
  \end{tabular}
\end{table}

ContentVec achieves the lowest residual (5.47\%), consistent with its explicit
disentanglement architecture. DPHuBERT (7.98\%) and Whisper-ppg (7.74\%) follow closely,
reflecting their content-focused design. WavLM Base+ (9.31\%) benefits from denoising
pre-training, while standard HuBERT variants without disentanglement objectives rank
highest in residual (14.23\%--18.82\%). A noteworthy layer-wise pattern emerges in
HuBERT LARGE: layer~18 carries 11.02\% residual, rising sharply to 18.82\% at layer~21.
Transformer layers 19--21 appear to accumulate speaker-specific information, which is
problematic because the downstream ASR model (HuBERT-Large-Finetuned) fine-tunes exactly
these layers while keeping earlier layers frozen.

\begin{figure}[t]
  \centering
  \begin{subfigure}[b]{0.235\textwidth}
    \centering
    \includegraphics[width=\textwidth]{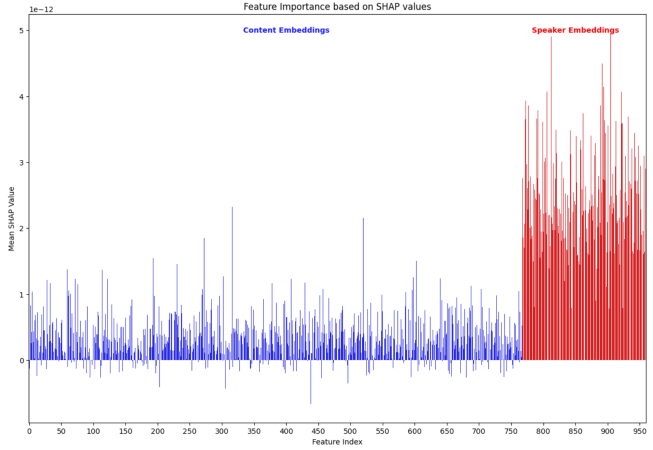}
    \caption{HuBERT BASE: speaker (orange) vs.\ content (blue) SHAP contributions.}
  \end{subfigure}
  \hfill
  \begin{subfigure}[b]{0.235\textwidth}
    \centering
    \includegraphics[width=\textwidth]{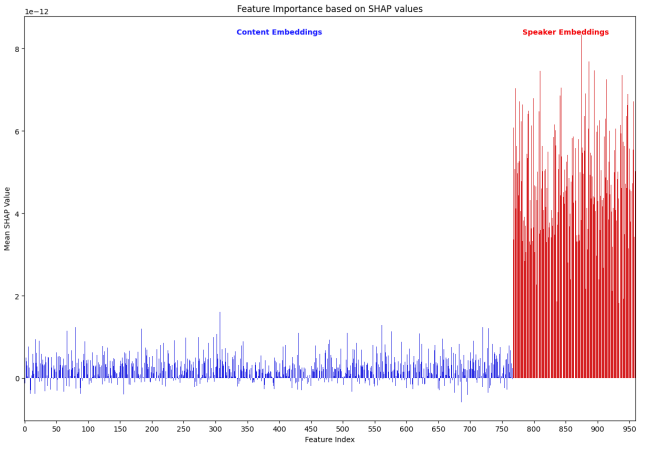}
    \caption{ContentVec: content (blue) attribution is much smaller than HuBERT BASE, consistent with 5.47\% residual.}
  \end{subfigure}
  \caption{SHAP attribution distributions. Each bar represents the mean absolute
    SHAP value of the content (blue) and speaker (orange) embedding segments. The
    content/speaker split visualises $R_M$ directly.}
  \label{fig:shap_dist}
\end{figure}

Figure~\ref{fig:shap_dist} visualises the mean absolute SHAP attribution distributions
for HuBERT BASE and ContentVec. In both models, the speaker embedding segment (orange)
makes the dominant positive contribution to the classification decision, as expected
given that $E_s$ is explicitly included in the concatenated input. Crucially, the
content segment (blue) is not attribution-free: both models exhibit a non-trivial
positive contribution from content dimensions, confirming that phonetic representations
carry residual timbre information. The key difference lies in \emph{magnitude}: HuBERT
BASE exhibits substantially larger content SHAP values, corresponding to its 14.23\%
residual in Table~\ref{tab:benchmark}, whereas ContentVec's content attribution bars are
considerably smaller, in line with its 5.47\% residual. This quantitative gap directly
reflects the effect of ContentVec's disentanglement objective, which suppresses
speaker-encoding content dimensions without driving their SHAP contributions to zero.

\subsection{Filtering results}
\label{ssec:filter_results}

Filtering experiments use HuBERT LARGE layer~21 embeddings---the worst-performing
configuration (18.82\% residual)---as input. For SHAP Noise, $\sigma \in \{-1.0,-0.6,-0.3\}$
is evaluated. For SHAP Cropping, $r=1\%$ (top 10 of 1024 dimensions) is fixed and
$\alpha \in \{0.99, 0.58\}$ is varied. Table~\ref{tab:filter} reports WER and
timbre residual after filtering. WER is computed on full ground-truth transcriptions
from the 40-speaker corpus.

\begin{table}[th]
  \caption{Speaker disentanglement filtering results on HuBERT LARGE layer~21
    embeddings. WER computed via seed-tts-eval. Lower residual and WER are better.}
  \label{tab:filter}
  \centering
  \resizebox{\columnwidth}{!}{%
  \begin{tabular}{llcc}
    \toprule
    \textbf{Method} & \textbf{Params} & \textbf{WER (\%)} & \textbf{Residual (\%)} \\
    \midrule
    Raw audio & ---       & 2.03 & --- \\
    No filter & ---       & 2.11 & 18.82 \\
    \midrule
    \multirow{3}{*}{SHAP Noise}
      & $\sigma=-1.0$ & 2.30 & \textbf{0.00} \\
      & $\sigma=-0.6$ & 2.16 & 2.21 \\
      & $\sigma=-0.3$ & 2.13 & 7.23 \\
    \midrule
    \multirow{2}{*}{SHAP Cropping}
      & $\alpha=0.99$ & 2.23 & 4.65 \\
      & $\alpha=0.58$ & 2.18 & 10.07 \\
    \midrule
    ContentVec (ref.) & L12 & 2.09 & 5.47 \\
    \bottomrule
  \end{tabular}
  }
\end{table}

SHAP Noise at $\sigma=-1.0$ eliminates speaker residuals entirely, at the cost of a
0.19 percentage point WER increase (2.30\% vs.\ 2.11\% baseline). The operating point
$\sigma=-0.6$ provides the most practical trade-off: residual drops to 2.21\% while WER
increases by only 0.05 percentage points. Notably, this surpasses ContentVec's 5.47\%
residual without any model retraining, demonstrating that post-hoc SHAP filtering can
exceed the disentanglement achieved by purpose-built training objectives.

SHAP Cropping reduces residuals to 4.65\% at best ($\alpha=0.99$), comparable to but
slightly worse than ContentVec, and with a larger WER cost than SHAP Noise at similar
residual levels. The performance gap between the two filters is examined below.

\subsection{Analysis}
\label{ssec:analysis}

\smallskip\noindent\textbf{Why SHAP Noise outperforms SHAP Cropping.} Timbre perception depends on
relative intensity and phase relationships across frequency components, encoded by both
positive and negative SHAP contributions. SHAP Cropping with $r=1\%$ suppresses only
the top-10 speaker-encoding dimensions (out of 1024), addressing a single tail of the
attribution distribution. SHAP Noise, by modulating \emph{all} dimensions in proportion
to their signed $\hat{\phi}_c$ values (Eq.~\ref{eq:noise_phi}--\ref{eq:noise_filter}),
disrupts the full speaker-encoding subspace, explaining its ability to drive residuals to
0\%.

\begin{figure}[t]
  \centering
  \includegraphics[width=0.46\textwidth]{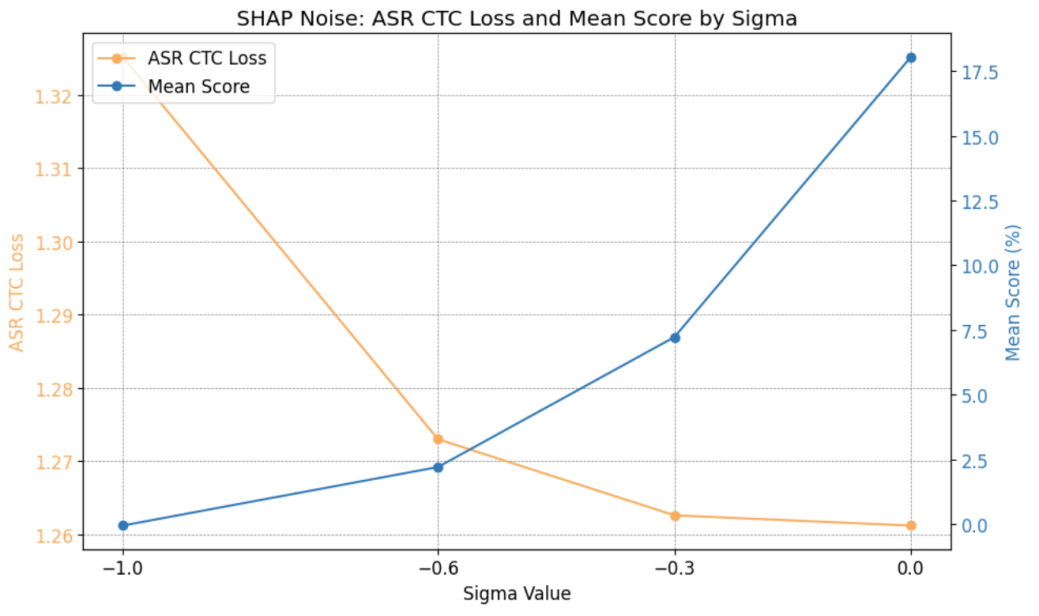}
  \caption{Trade-off between speaker residual (Mean Score, left axis) and content
    preservation (CTC Loss, right axis) as a function of $\sigma$ for SHAP Noise
    filtering. CTC loss is shown here as a granular local diagnostic; global content
    quality is reported as WER in Table~\ref{tab:filter}.}
  \label{fig:param_analysis}
\end{figure}

\smallskip\noindent\textbf{Parameter sensitivity.} Figure~\ref{fig:param_analysis} shows the trade-off
between CTC loss and speaker residual as $\sigma$ varies. Residual decreases
exponentially with $|\sigma|$, while CTC loss rises sharply only beyond $\sigma=-0.8$.
The plateau around $\sigma=-0.6$ defines the practical operating region, achieving
88.3\% residual reduction relative to the unfiltered baseline while increasing CTC loss
by less than 1\%.

\smallskip\noindent\textbf{Layer sensitivity.} Layer~18 embeddings of HuBERT LARGE yield a WER of 2.09\%
through the HuBERT-Large-Finetuned ASR model, nearly matching the raw audio baseline of
2.03\%, which confirms that the ASR fine-tuning left earlier layers unchanged. Layer~21
embeddings yield 2.11\% WER but carry 18.82\% residual---substantially higher than
layer~18's 11.02\%---because ASR fine-tuning layers 18--21 acquire task-relevant
representations that incidentally encode speaker cues. From a disentanglement standpoint,
layer~18 is the preferable extraction target for content-centric downstream tasks.

\section{Conclusion}

We have shown that SHAP-based interpretability provides both a direct measurement and an
effective remedy for speaker entanglement in speech pre-trained models. The
InterpTRQE-SptME benchmark quantifies timbre residual directly, rather than through
indirect downstream proxies, and works without any model modification. InterpTF-SptME
filtering---particularly the SHAP Noise variant---reduces
speaker residuals from 18.82\% to near zero while keeping WER within 2.3\%, and
surpasses the disentanglement level of ContentVec without retraining. The approach is
model-agnostic, layer-agnostic, and label-free, making it applicable to any existing
speech representation system. Future work should explore extending the benchmark to
multilingual scenarios and applying SHAP-guided filtering to other latent attributes
beyond speaker identity, such as emotional prosody and channel conditions.

\bibliographystyle{IEEEtran}
\bibliography{mybib}

\end{document}